# Lattice Vibration, Raman Modes and Room-Temperature Spin-Phonon Coupling in Intrinsic 2D van der Waals Ferromagnetic $Fe_3GaTe_2$

Gaojie Zhang[1,2], Hao Wu[1,2], Li Yang[1,2], Wen Jin[1,2], Bichen Xiao[1,2], Wenfeng Zhang[1,2,3], Haixin Chang[1,2,3,*]

[1]State Key Laboratory of Material Processing and Die & Mold Technology, School of Materials Science and Engineering, Huazhong University of Science and Technology, Wuhan 430074, China.

[2]Wuhan National High Magnetic Field Center and Institute for Quantum Science and Engineering, Huazhong University of Science and Technology, Wuhan 430074, China.

[3]Shenzhen R&D Center of Huazhong University of Science and Technology, Shenzhen 518000, China.

[*]Corresponding authors. E-mail: hxchang@hust.edu.cn

Two-dimensional (2D) van der Waals (vdW) magnets with spin-phonon coupling are crucial for next-generation spintronics. Among them, $Fe_3GaTe_2$ has attracted widespread attention due to above-room-temperature intrinsic ferromagnetism and large perpendicular magnetic anisotropy. However, the lattice vibrations and the interplay between ferromagnetism and lattice vibrations in $Fe_3GaTe_2$ are still unexplored. Here, we report the lattice vibration, Raman modes, and room-temperature spin-phonon coupling in 2D $Fe_3GaTe_2$ with above-room-temperature Curie temperature ($T_C$). Two typical Raman modes with out-of-plane lattice vibrations are identified: $A_{1g}^1$ and $A_{1g}^2$, whose frequencies increase as the thickness decreases from bulk to 2D $Fe_3GaTe_2$ due to the weakening of interlayer vdW interactions and spin exchange coupling. Moreover, the difference between phonon band dispersions under ferromagnetic and nonmagnetic interlayer spin ordering indicates the existence of spin-phonon coupling. The phonon frequency diverges from the anharmonic model below $T_C$ and thus the strength of spin-phonon coupling is ~0.28 $cm^{-1}$ at 300 K, which is the first experimental identification of room-temperature spin-phonon coupling in 2D vdW magnets. This work deepens the understanding of novel 2D vdW magnets and provides a basis for spintronic applications at and above room temperature.

## INTRODUCTION

Two-dimensional (2D) materials with van der Waals (vdW) structures have become the foundation for fundamental science and technological application in next-generation electronics, photonics, and spintronics[1-5]. Among them, the first discovery of intrinsic 2D magnetism in 2D atomic crystals $CrI_3$[6] and $Cr_2Ge_2Te_6$[7] refreshed the inherent understanding of Mermin-Wagner theorem[8]. Motivated by these insulating chromium compounds, intrinsic 2D ferromagnetism was subsequently discovered in metallic $Fe_nGeTe_2$ (n = 3, 4, 5)[9, 10] and $CrTe_2$[11], offering an unprecedented opportunity for manipulating the spin and charge degrees of freedom in the 2D limit. Exploring the interplay between 2D magnetism and energy quanta of lattice vibrations is critical for understanding various novel phenomena and developing new spin physics. Meanwhile, spin-phonon coupling plays an important role in many physical phenomena, such as phonon Hall effect[12], spin-Seebeck effect[13], and spin-Peierls transition[14]. However, previous studies on spin-phonon coupling have been conducted in 2D vdW magnets at rather low cryogenic temperature[14-18]. No room-temperature spin-phonon coupling is reported in 2D vdW magnets so far.

The recently-discovered 2D vdW ferromagnetic crystal $Fe_3GaTe_2$ has generated tremendous interest due to its intrinsic ferromagnetism above room temperature (Curie temperature $T_C \approx$ 350-380 K) and large perpendicular magnetic anisotropy[19], thereby leading to a series of all-vdW-integrated room-temperature spintronic devices[20-22]. $Fe_3GaTe_2$ also exhibits a wide range of interesting physics phenomena up to room temperature, such as highly-tunable coercivity[23], topological nodal lines[24], field-free domain modulation[25], magnetic skyrmions[26], and giant topological Hall effect[27]. However, the lattice vibrations, Raman modes, and spin-phonon coupling remain largely unexplored for $Fe_3GaTe_2$.

Here, we study the lattice vibrations, Raman modes, and spin-phonon coupling in $Fe_3GaTe_2$ through first-principles calculations and systematic Raman measurements. Two prominent Raman modes with out-of-plane lattice vibrations, $A_{1g}^1$ and $A_{1g}^2$, are observed at room temperature in 2D $Fe_3GaTe_2$ with thickness ranging from 143 to 8 nm. Moreover, the comparison of phonon band dispersions under ferromagnetic and nonmagnetic interlayer spin ordering suggests the existence of spin-phonon coupling. The temperature-dependent phonon frequency deviates from the anharmonic model below $T_C$ and thus confirms the strength of spin-phonon coupling of ~0.28 cm$^{-1}$ at 300 K, which is the first experimental identification of room-temperature spin-phonon coupling in 2D vdW magnets. These results lay a foundation for the ferromagnetism and lattice vibrations at room temperature or above for 2D vdW magnets.

## RESULTS AND DISCUSSION

**Structural and ferromagnetic properties of $Fe_3GaTe_2$.** High-quality $Fe_3GaTe_2$ single crystals are grown via the chemical vapor transport (CVT) method (see details in the **Experimental Section**). As shown in **Figure 1a**, the $Fe_3GaTe_2$ shows the vdW-stacked structure along c axis with a $Fe_3Ga$ heterometallic layers sandwiched between two Te layers. Two types of Fe atoms (labeled $Fe_I$ and $Fe_{II}$) with different local chemical environments can be identified in the crystal lattice. The phase of as-grown bulk $Fe_3GaTe_2$ single crystals is confirmed by the X-ray diffraction (XRD) pattern, showing a series of diffraction peaks belonging to the (00$l$) ($l$ is even) crystal plane (**Figure 1b**). This result implies a strict growth orientation of the crystal along the c axis, which is consistent with previous works[19, 28]. According to the Bragg's law, the interlayer distance is calculated as ~0.8 nm. Moreover, the atomic arrangements and vdW structure of as-grown $Fe_3GaTe_2$ are further confirmed by high-angle annual dark-field scanning transmission electron microscopy (HAADF-STEM) along the [100] orientation, as depicted in **Figure 1c**. The clear atomic arrangements and interlayer distance match perfectly with the standard $Fe_3GaTe_2$ crystal structure and XRD results, reconfirming the high quality of as-grown crystals. Meanwhile, by using the energy-dispersive X-ray spectroscopy (EDS), the atomic ratio of Fe, Ga, and Te is determined to be 50.75:16.56:32.69, very close to ideal 3:1:2 stoichiometry (**Figure S1**).

Due to the strong perpendicular magnetic anisotropy of $Fe_3GaTe_2$, the intrinsic magnetic properties of bulk $Fe_3GaTe_2$ crystals are measured under the out-of-plane magnetic fields. As shown in **Figure 1d**, temperature-dependent magnetization (M-T) curves with zero-field-cooling (ZFC) and field-cooling (FC) protocols (B = 0.1 T) demonstrate the ferromagnetic feature with a sharply-weakened magnetization around $T_C$. By taking the first derivation (dM/dT), the $T_C$ of the bulk $Fe_3GaTe_2$ crystals is determined to be ~366 K (inset in **Figure 1d**), higher than other known vdW intrinsic ferromagnets[6, 7, 9, 29]. Meanwhile, isothermal magnetization (M-B) curves with a series of hysteresis loops reveal the intrinsic ferromagnetism above room temperature (**Figure 1e**). It should be noted that this room-temperature saturation magnetization ($M_S$) is as high as 48.8 emu/g, which is 3.9 times that of another room-temperature vdW ferromagnetic crystal CrTe$_2$ ($M_S \approx 12.5$ emu/g)[29]. The room-temperature long-range ferromagnetic order of bulk Fe$_3$GaTe$_2$ is further confirmed by an obvious anomalous Hall effect with $R_{AH} \approx 0.002$ Ω at 300 K (**Figure 1f**).

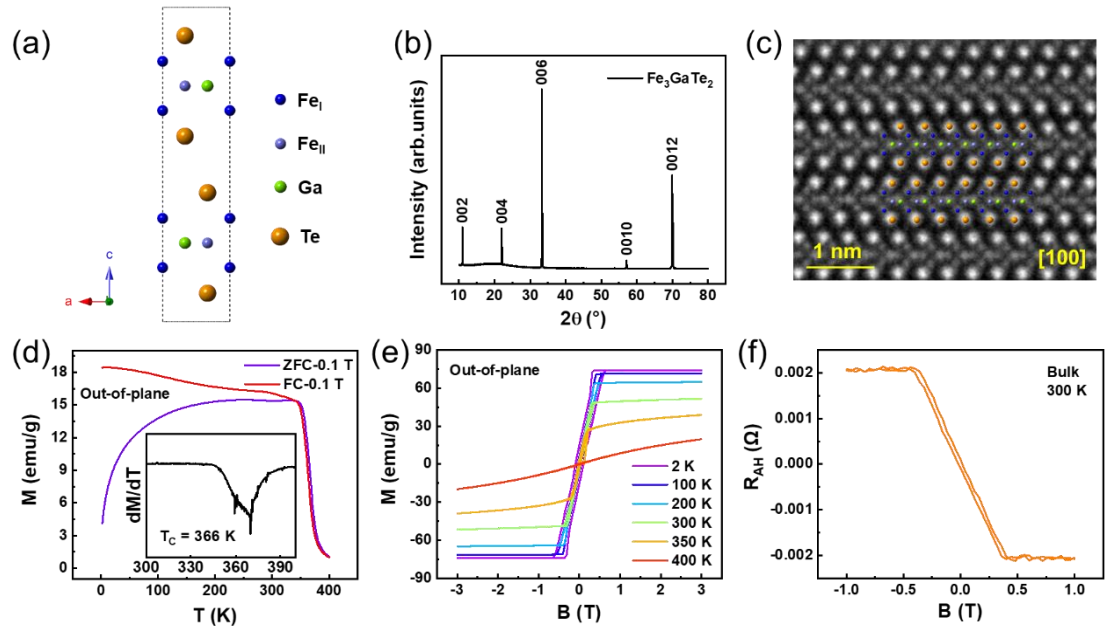

**Figure 1. Structural and magnetic characterizations of vdW Fe$_3$GaTe$_2$ crystals.** (a) Schematic of crystal structure of Fe$_3$GaTe$_2$. (b) XRD pattern of bulk Fe$_3$GaTe$_2$. (c) Cross-sectional atomic-resolution HAADF-STEM image of Fe$_3$GaTe$_2$. Inset shows the atomic model that matches well with the image. (d) ZFC-FC curves of bulk Fe$_3$GaTe$_2$ under out-of-plane magnetic field of 0.1 T. Inset is the first derivation of the ZFC curve. (e) Isothermal magnetization curves of bulk Fe$_3$GaTe$_2$ under out-of-plane magnetic field and different temperatures. (f) Anomalous Hall effect of bulk Fe$_3$GaTe$_2$ at 300 K.

Next, to further identify the above-room-temperature intrinsic ferromagnetism in 2D $Fe_3GaTe_2$, the magneto-transport measurement is conducted on the Hall device based on an exfoliated $Fe_3GaTe_2$ nanosheet from the bulk single crystal (**Figure 2a**). The thickness of this as-tested $Fe_3GaTe_2$ is unveiled by AFM test, indicating the thickness is 13 nm (**Figure 2b**). As shown in **Figure 2c**, a series of clear hysteresis loops indicates the anomalous Hall effect, suggesting the above-room-temperature long-range intrinsic ferromagnetism in this 2D $Fe_3GaTe_2$. Meanwhile, compared with the bulk value of ~366 K, its $T_C$ shows a slight decrease to ~350 K, which is common in vdW magnets when their thickness down to the 2D regime[6, 7, 9]. These results make $Fe_3GaTe_2$ a unique candidate for studying lattice vibrations, Raman modes, and spin-phonon coupling of 2D vdW magnets at and above room temperature.

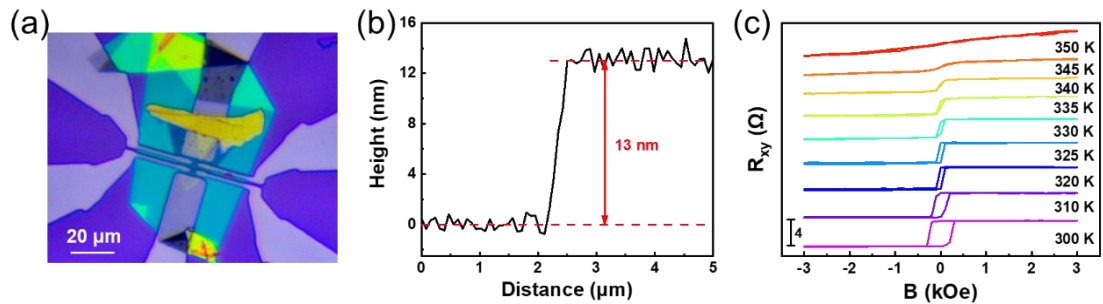

**Figure 2. Above-room-temperature intrinsic ferromagnetism in 2D $Fe_3GaTe_2$.** (a) Optical image of the Hall device based on the 2D $Fe_3GaTe_2$ nanosheet. (b) AFM profile height curve of the as-tested 13 nm $Fe_3GaTe_2$ nanosheet. (c) Magnetic field-dependent Hall resistance ($R_{xy}$-B) curves under different temperatures above room temperature.

**Lattice vibrations and thickness-dependent Raman modes of $Fe_3GaTe_2$.** Raman spectroscopy is a fast and non-destructive method for studying lattice vibrations. In this work, the evolution of lattice vibrations and Raman modes with $Fe_3GaTe_2$ thickness is investigated by combining first-principles calculations and Raman measurements (see details in the **Experimental Section**). Based on first-principles calculations, **Figure 3a** shows the atomic displacement patterns of two typical Raman modes ($A_{1g}^1$ and $A_{1g}^2$) for $Fe_3GaTe_2$. The $A_{1g}^1$ mode indicates the out-of-plane vibrations of the $Fe_I$, $Fe_{II}$, Ga, and Te atoms with the vibration amplitude Ga>Te>$Fe_{II}$>$Fe_I$, while the $A_{1g}^2$ mode indicates the out-of-plane vibrations of the $Fe_I$ and Te atoms with the vibration amplitude Te>$Fe_I$. With the reduction of the thickness, the frequency ($\omega_p$) of these two Raman modes increases (**Figure 3b**). For the $A_{1g}^1$ mode, the $\omega_p$ increases from 107 $cm^{-1}$ in the bulk to 109.3 $cm^{-1}$ in the bilayer to 109.5 $cm^{-1}$ in the monolayer. For the $A_{1g}^2$ mode, the $\omega_p$ increases from 125.4 $cm^{-1}$ in the bulk to 127 $cm^{-1}$ in the bilayer to 132.1 $cm^{-1}$ in the monolayer. It is interesting that the frequency change ($\Delta\omega_p$) of $A_{1g}^2$ mode from the bulk to monolayer ($\Delta\omega_p \approx 6.7$ $cm^{-1}$) is larger than that of some other 2D materials. For example, graphene and $MoS_2$ exhibit the relatively small $\Delta\omega_p$ of ~3-5 $cm^{-1}$ from bulk to 2D, which mainly originates from the slightly reduced interlayer coupling strength dominated by reduced weak vdW interactions[30-32]. By contrast, a comparably large $\Delta\omega_p$ of ~5-9 $cm^{-1}$ is also observed in another 2D material $PdSe_2$ with strong interlayer coupling, which is attributed to the largely reduced interlayer coupling caused by reduced vdW interactions and interlayer covalent hybridizations[33]. Such reduced covalent hybridization can be implied by the in-plane lattice shrinking effect with decreasing thickness. However, for $Fe_3GaTe_2$, the decrease of in-plane lattice constants from bulk to monolayer is only ~0.17%, indicating a very weak lattice shrinking effect (**Table S1**). Meanwhile, it should be noted that the difference between Fe$_3$GaTe$_2$ and the aforementioned 2D materials is that the former hosts strong intrinsic ferromagnetism in each layer, and thus the interlayer spin exchange coupling contributes to additional interlayer coupling[34]. Therefore, the large $\Delta\omega_p$ in intrinsic ferromagnetic Fe$_3$GaTe$_2$ can be ascribed to the strong interlayer coupling mainly made of vdW interactions and spin exchange coupling between the layers.

Then, the experimental Raman measurements are conducted on hBN-encapsulated Fe$_3$GaTe$_2$ nanosheets with a thickness range of 143 to 8 nm (**Figure S2**). According to the Raman spectra, the $A_{1g}^1$ and $A_{1g}^2$ modes with $\omega_p$ of ~103.2-105 cm$^{-1}$ and ~125.5-126.6 cm$^{-1}$ can be observed at 300 K (**Figure 3c**). The $\omega_p$ of $A_{1g}^1$ and $A_{1g}^2$ modes are extracted and show a certain thickness dependence (**Figure 3d**). As the thickness decreases, $\omega_p$ gradually increases, consistent with the theoretical evolution of $\omega_p$ with thickness shown in **Figure 3b**. Meanwhile, it is worth noting that the $\omega_p$ increases rapidly when the thickness of Fe$_3$GaTe$_2$ is less than 17.3 nm, which can be ascribed to the change of strong interlayer coupling with the thickness and the more significant substrate effect in the 2D regime[34, 35].

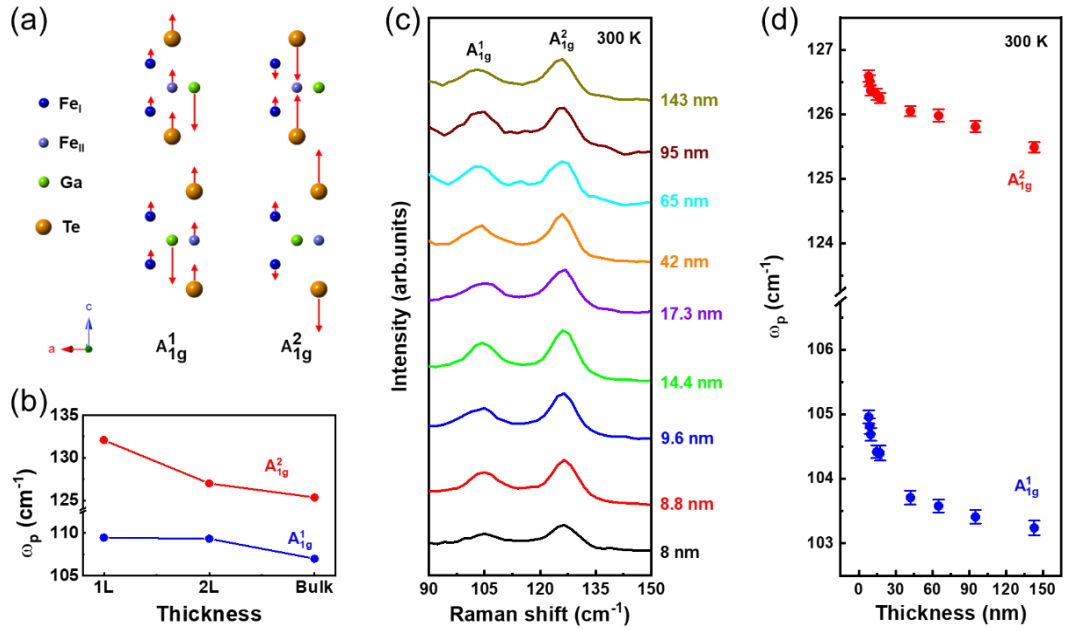

**Figure 3. Lattice vibration and thickness-dependent Raman modes in Fe₃GaTe₂ from first-principles calculations and room-temperature Raman tests.** (a) Atomic displacement patterns of Fe₃GaTe₂ for $A_{1g}^1$ and $A_{1g}^2$ modes. The interlayer spin ordering is ferromagnetic. The length of the red arrow represents the relative vibration amplitude of each atom. (b) The evolution of the theoretical frequencies for $A_{1g}^1$ and $A_{1g}^2$ modes with the Fe₃GaTe₂ thickness. (c) Experimental Raman spectra of 2D Fe₃GaTe₂ nanosheets with different thicknesses at 300 K. (d) Extracted frequencies of $A_{1g}^1$ and $A_{1g}^2$ modes at different Fe₃GaTe₂ thicknesses. Error bar SD, N=3.

Most 2D magnets are sensitive to water and oxygen in the atmospheric environment[6, 7, 36], and Raman spectroscopy could potentially be used to identify oxidation situation of the sample. Therefore, the Raman spectra of naturally oxidized Fe₃GaTe₂ has also been studied. **Figure 4a** shows the optical image of Fe₃GaTe₂ nanosheets without the hBN encapsulation. According to the AFM test, the thicknesses of them are 12 and 44 nm, respectively (**Figure 4b**). The Raman spectra of oxidized Fe₃GaTe₂ nanosheets are totally different from those of the hBN-encapsulated pristine ones. Specifically, the $A_{1g}^1$ and $A_{1g}^2$ modes disappear, and two new Raman peaks emerge, located at ~119.3 and ~138.9 cm$^{-1}$ for 44 nm Fe$_3$GaTe$_2$ and ~122.1 and ~140.8 cm$^{-1}$ for 12 nm Fe$_3$GaTe$_2$, respectively (**Figure 4c**). Raman peaks with similar locations can also be observed in another isomorphic 2D vdW ferromagnet Fe$_3$GeTe$_2$ after natural oxidation[37]. Therefore, such two new Raman peaks in naturally oxidized Fe$_3$GaTe$_2$ imply the degradation of the crystal after exposing in the air condition[15].

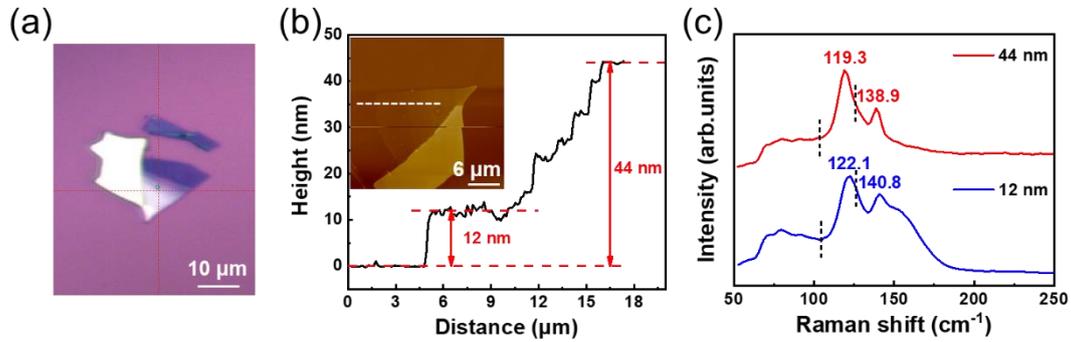

**Figure 4. Room-temperature Raman measurements of naturally-oxidized Fe$_3$GaTe$_2$ nanosheets.** (a) Optical image of the as-tested Fe$_3$GaTe$_2$ nanosheets without the hBN encapsulation. (b) AFM image and corresponding profile height curve along the white dash line. (c) Raman spectra of the 12 and 44 nm naturally-oxidized Fe$_3$GaTe$_2$ nanosheets. For comparison, black dash lines indicate the locations of two Raman peaks in hBN-encapsulated pristine Fe$_3$GaTe$_2$ with the similar thickness (i.e. 14.4 and 42 nm).

**Spin-phonon coupling in Fe$_3$GaTe$_2$.** To study the effect of spin-phonon coupling on the lattice vibration, the first-principles calculations are performed on bulk, bilayer, and monolayer Fe$_3$GaTe$_2$ with either ferromagnetic or nonmagnetic interlayer spin ordering (see details in the **Experimental Section**). As depicted in **Figure 5a-c**, some important features emerge. First, since there are no negative phonon branches, both ferromagnetic and nonmagnetic ordering are mechanically stable in $Fe_3GaTe_2$ from bulk to monolayer[38]. Second, a small phonon band gap located in 6 to 8 THz can be identified in $Fe_3GaTe_2$ from bulk to monolayer. All the ferromagnetic $Fe_3GaTe_2$ have narrower phonon band gaps than the corresponding nonmagnetic ones, implying the spin-induced narrowing of phonon band gaps. Third, compared to the nonmagnetic cases, bulk, bilayer, and monolayer ferromagnetic $Fe_3GaTe_2$ exhibit visible shifts of both optical and acoustic phonon branches towards lower phonon frequencies. This feature indicates the existence of spin-phonon coupling in $Fe_3GaTe_2$[38, 39].

Further, **Figure 5d-f** present the corresponding total and partial phonon density of states (PDOS) of $Fe_3GaTe_2$ with either ferromagnetic or nonmagnetic interlayer spin ordering. The relative atomic masses of Fe, Ga, and Te are 55.845, 69.723, and 127.6, respectively. In general, heavy atoms produce low-frequency vibrations, while light atoms produce high-frequency vibrations[38]. Therefore, the total PDOS of the low-frequency modes (e.g. <4 THz for ferromagnetic $Fe_3GaTe_2$) mainly consist of vibrations of Te atoms, while the high-frequency modes (e.g. >4 THz for ferromagnetic $Fe_3GaTe_2$) mainly come from the vibrations of Fe atoms, as presented in **Figure 5d-f**. By contrast, due to the moderate relative atomic mass and minimal quantity of the Ga atom in $Fe_3GaTe_2$, its mainly contributes to the total PDOS at very few frequencies. In addition, compared to the nonmagnetic cases, obvious shifts towards lower phonon frequencies of the total,

Fe, and Te partial PDOS can also be observed in ferromagnetic $Fe_3GaTe_2$, reconfirming the existence of spin-phonon coupling.

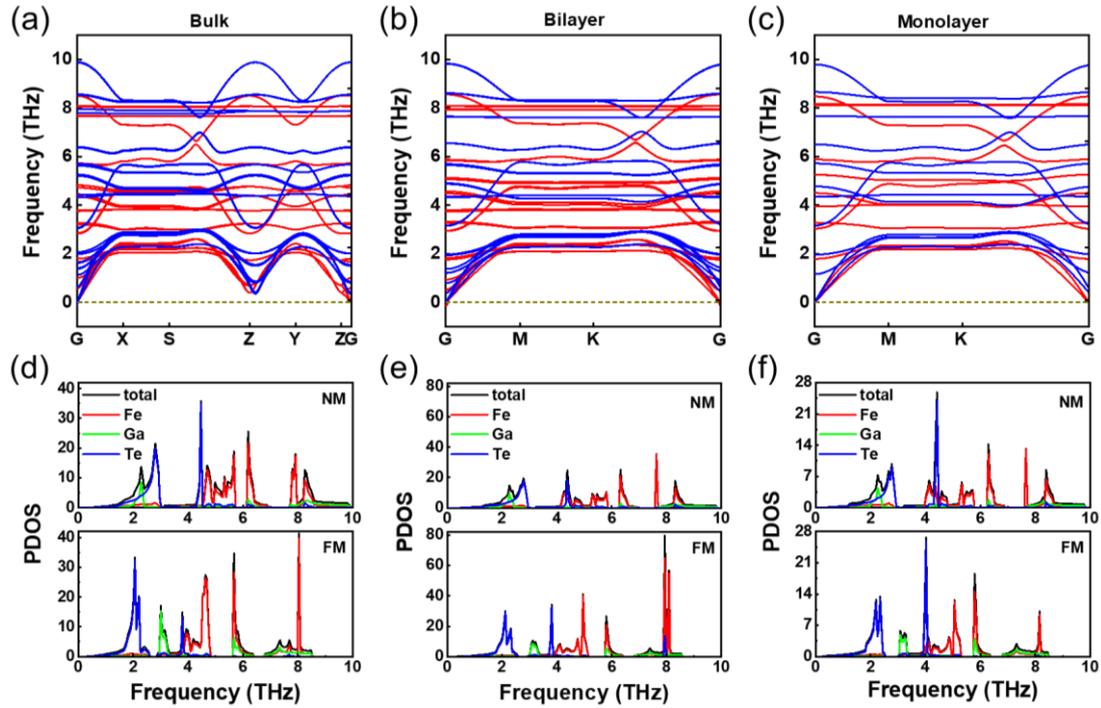

**Figure 5. Phonon band dispersions and phonon density of states (PDOS) for $Fe_3GaTe_2$ under ferromagnetic (FM) and nonmagnetic (NM) interlayer spin ordering.** (a-c) Phonon band dispersions of bulk, bilayer, and monolayer $Fe_3GaTe_2$. The phonon dispersions of ferromagnetic and nonmagnetic are depicted by red and blue lines, respectively. (d-f) PDOS of bulk, bilayer, and monolayer $Fe_3GaTe_2$.

After theoretical discussion of spin-phonon coupling in $Fe_3GaTe_2$, magneto-transport and Raman measurements are conducted at varying temperatures. **Figure 6a** exhibits the optical image and profile height curve of an as-tested 47 nm $Fe_3GaTe_2$ nanosheet. The temperature-dependent longitudinal resistance ($R_{xx}$-T) curve shows a typical metallic nature, where the $R_{xx}$ gradually decreases as the temperature decreases (**Figure**

**6b**). Due to the second-order nature of magnetic phase transition, a kink can be observed around 360 K in the $R_{xx}$-T curve, corresponding to the $T_C$ of this 47 nm Fe$_3$GaTe$_2$ nanosheet. Meanwhile, the above-room-temperature intrinsic ferromagnetism is also confirmed by a series of anomalous Hall effect curves ranging from 300 to 360 K (**Figure 6c**). When the applied magnetic field sweeps from negative to positive, the $R_{xy}$ first jumps sharply and then increases slowly. This two-step behavior is attributed to the multi-domain structure, which has also been observed in other 2D ferromagnets[40].

Further, temperature-dependent Raman measurements from 300 to 400 K are conducted on this 47 nm Fe$_3$GaTe$_2$ nanosheet. Two distinct Raman modes, including $A_{1g}^1$ (~105 cm$^{-1}$) and $A_{1g}^2$ (~127 cm$^{-1}$), can be observed in all temperatures ranging from 300 to 400 K (**Figure 6d**). Since $A_{1g}^1$ mode possesses larger linewidths and lower intensities, we focus on the $A_{1g}^2$ mode to study the spin-phonon coupling. The temperature-driven evolution of phonon frequency ($\omega_p$) for the $A_{1g}^2$ mode are extracted in **Figure 6e**, which can be analyzed by the anharmonic model[15, 41, 42]:

$$\omega_{anhp}(T) = \omega_p(0) - A\left[1 + \frac{2}{e^{\frac{\hbar\omega_p(0)}{2k_BT}} - 1}\right]$$

where $\omega_p(0)$ is the phonon frequency at zero temperature, A the fitting parameter, $\hbar$ the reduced Planck constant, and $k_B$ the Boltzmann constant. As depicted in **Figure 6e**, as the temperature decreases from 400 to 300 K, the $\omega_p$ gradually increases and deviates from the standard anharmonic model (red line) with lower phonon frequencies. The deviation of $\omega_p$ from the standard anharmonic model, defined as $\Delta\omega_p(T) = \omega_p(T) -$

$\omega_{anhp}(T)$, is presented in **Figure 6f**. It turns out that the $\omega_p$ reduces quickly below 360 K, consistent with the $T_C$ of this 47 nm $Fe_3GaTe_2$ nanosheet. This result suggests the existence of spin-phonon coupling below $T_C$, which may originate from the modulation effect of the super-exchange integral by lattice vibrations[43, 44]. In a nearest-neighbor approximation, the $\Delta\omega_p$ can be further expressed as $\lambda <S_i \cdot S_j>$, where $\lambda$ is the strength of the spin-phonon coupling and $<S_i \cdot S_j>$ is the spin correlation function of nearest-neighbor spins[14, 45, 46]. For $Fe_3GaTe_2$, the $M_S$ is ~48.8 emu/g ($\approx 1.43$ μB/Fe) at 300 K (**Figure 1e**) and thus the corresponding $<S_i \cdot S_j>$ is ~0.51[14]. Since the $\Delta\omega_p$ is ~0.143 cm$^{-1}$ at 300 K (**Figure 6f**), the strength of room-temperature spin-phonon coupling $\lambda \approx 0.28$ cm$^{-1}$, indicating the first experimental identification of such coupling effect at room temperature in 2D vdW magnets. For comparison, previous studies about the strength of spin-phonon coupling in 2D vdW magnets are summarized in **Table 1**, all of which could only be conducted at rather low temperatures. Therefore, the observed results in 2D $Fe_3GaTe_2$ is crucial for further understanding potential physical phenomena and spintronic devices based on 2D magnets at or even above room temperature.

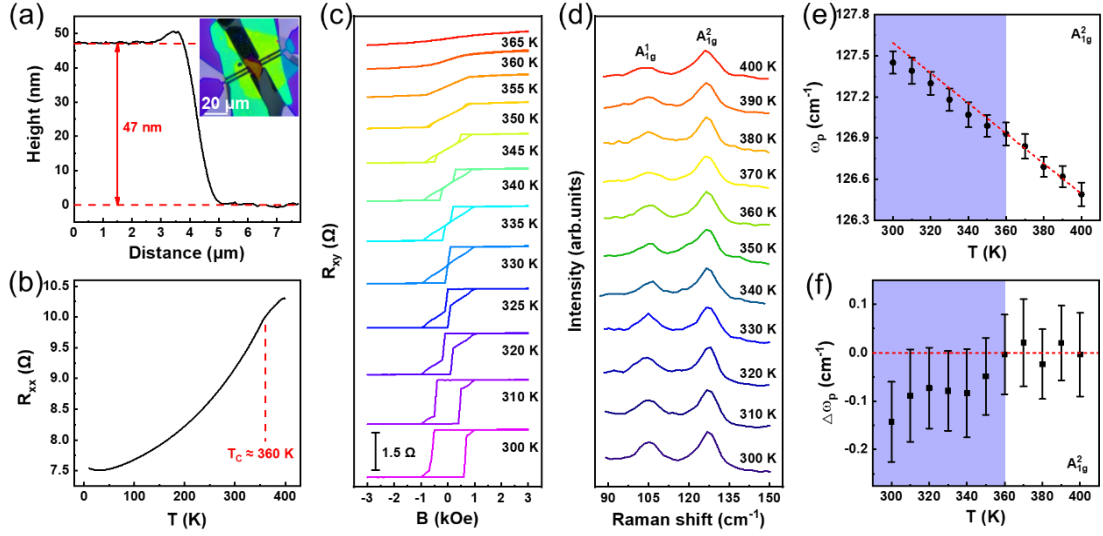

**Figure 6. Magneto-transport measurements and spin-phonon coupling in Fe₃GaTe₂ nanosheet.** (a) Optical image and AFM profile height curve of the as-tested Fe₃GaTe₂-based Hall device. (b) Temperature-dependent longitudinal resistance ($R_{xx}$-T) curve. The kink shows the ferromagnetic-paramagnetic transition, and the $T_C$ is ~360 K. (c) Magnetic field-dependent Hall resistance ($R_{xy}$-B) curves under different temperatures. (d) Raman spectra under different temperatures. (e) Extracted phonon frequencies of $A_{1g}^2$ modes at different temperatures. The red dash curve is a fitting curve based on the anharmonic model. Error bar SD, N=3. (f) The deviation of phonon frequency from the standard anharmonic model at different temperatures. The red dash curve is set to $\Delta\omega_p = 0$. Error bar SD, N=3.

**Table 1. Summary of the typical strength of spin-phonon coupling in 2D vdW magnets.**

| 2D vdW magnets | T (K) | λ (cm⁻¹) | References |
| --- | --- | --- | --- |
| Cr₂Ge₂Te₆ | 10 | 0.24-1.2 | 14 |
| Fe₃GeTe₂ | 30 | 1.36 | 15 |
| CrBr₃ | 10 | 0.31-0.51 | 17 |
| | 19 | 0.13-0.4 | 16 |
| CrSiTe₃ | 10 | 0.1-0.2 | 18 |
| Fe$_3$GaTe$_2$ | 300 | 0.28 | This work |

## CONCLUSIONS

In conclusion, experimental and theoretical investigations of lattice vibration, Raman modes, and room-temperature spin-phonon coupling are conducted on 2D ferromagnetic crystal Fe$_3$GaTe$_2$ with T$_C$ up to 366 K. Two Raman modes with out-of-plane vibrations are observed: $A_{1g}^1$ and $A_{1g}^2$, whose phonon frequencies increase with decreasing thickness and temperature. In addition, spin-phonon coupling is revealed through the phonon band dispersions and the deviation of anharmonic model below T$_C$. At 300 K, the strength of spin-phonon coupling is inferred to be ~0.28 cm$^{-1}$, achieving the first experimental identification of such coupling effect at room temperature for 2D vdW magnets. This finding deepens the understanding of novel physical properties and promotes the room-temperature applications of ultra-compact spintronics based on vdW magnets.

## EXPERIMENTAL SECTION

**Single crystal growth.** The Fe$_3$GaTe$_2$ crystals were grown using the chemical vapor transport (CVT) method. Specifically, Fe powder (99.95%, Aladdin), GaTe powder (99.99%, Aladdin), and Te powder (99.99%, Aladdin) with a molar ratio of 3:1:1 were mixed. Then, the mixture and a certain amount of I$_2$ granules (99.99%, Aladdin) were placed in a quartz ampoule, followed by evacuating and sealing. The ampoule was placed in a two-zone furnace. The temperatures of the source and growth zones were

750-800°C and 700-750°C, respectively, and the reaction maintained 336 hours. Finally, the furnace was naturally cooled to room temperature.

**Structural and thickness measurements.** The structure was tested by powder X-ray diffractometer (XRD, XRD-7000, SHIMADZU) with the Cu K$\alpha$ radiation ($\lambda$ = 0.154 nm) and aberration corrected transmission electron microscope (ACTEM, Themis Z, Thermo Fisher Scientific) with the energy-dispersive X-ray spectrometer (EDS). The [100]-orientated lamellae sample for the ACTEM test was fabricated using a focused ion beam (FIB) technique. The thickness was tested by atomic force microscope (AFM, Dimension EDGE, Bruker). All these tests were done at room temperature.

**Magnetic and electrical measurements.** The magnetization and magneto-transport tests were conducted by physical property measurement system (PPMS, DynaCool, Quantum Design) with vibrating sample magnetometer (VSM) and constant current modes, respectively. The currents applied to the bulk and nanosheet samples were 100 and 30 μA, respectively. The magnetic field was perpendicular to the 2D plane of the sample.

**Raman measurements.** The Raman spectra were obtained by laser confocal Raman spectrometer (Raman, LabRAM HR800, Horiba Jobin-Yvon) with an excitation wavelength of 532 nm and a laser spot of 1 μm. The sample was placed in a thermostat and the temperature was adjusted through a temperature control system. To stabilize the temperature, wait for 15 minutes at each temperature before testing.

**First-principles calculations.** The density functional theory (DFT)-based first-principles calculations[47, 48] were conducted in the Vienna ab initio simulation package (VASP). The plane-wave basis sets was treated by the projector augmented-wave method[49, 50]. The exchange-correlation potential was treated by a generalized gradient approximation (GGA) with the Perdew-Burke-Ernzerhof (PBE) parametrization[51]. The GGA+U method was used to treat the partially occupied 3d electrons of Fe element, wherein U = 1.5 eV[28, 52]. The interlayer vdW interactions were treated by the optB86-vdW method[53, 54] and the energy cutoff was set to 520 eV. The Brillouin-zone integration was sampled with a Γ-centered Monkhorst-Pack mesh of 24 × 24 × 6 and 24 × 24 × 1 for bulk and bilayer/monolayer $Fe_3GaTe_2$ slab models, respectively[55], with a vacuum region of ~20 Å along the z-axis. The convergence criteria for energy and forces were $10^{-5}$ eV and 0.01 eV/Å, respectively. For phonon relative calculations, the finite difference method as implemented in the Phonopy software[56] was used under ferromagnetic or nonmagnetic interlayer spin ordering. The convergence criteria for energy and forces were $10^{-8}$ eV and 0.001 eV/Å, respectively. Atomic displacements of Raman modes at different frequencies were obtained using JMOL software.

## ASSOCIATED CONTENT

**Data Availability Statement**

The data that support the findings of this study are available from the corresponding author upon reasonable request.

**Supporting Information**

The Supporting Information is available free of charge.

EDS spectra of an as-grown bulk $Fe_3GaTe_2$ crystal; Optical images and AFM tests of hBN-encapsulated $Fe_3GaTe_2$ nanosheets for room-temperature Raman measurements; In-plane lattice constants of bulk, bilayer, and monolayer $Fe_3GaTe_2$ atomic models (PDF)

**AUTHOR INFORMATION**

**Corresponding Author**

**Haixin Chang** - *State Key Laboratory of Material Processing and Die & Mold Technology, School of Materials Science and Engineering, Huazhong University of Science and Technology, Wuhan 430074, China; Wuhan National High Magnetic Field Center and Institute for Quantum Science and Engineering, Huazhong University of Science and Technology, Wuhan 430074, China; Shenzhen R&D Center of Huazhong University of Science and Technology, Shenzhen 518000, China;* Email: hxchang@hust.edu.cn

**Authors**

**Gaojie Zhang** - *State Key Laboratory of Material Processing and Die & Mold Technology, School of Materials Science and Engineering, Huazhong University of*

*Science and Technology, Wuhan 430074, China; Wuhan National High Magnetic Field Center and Institute for Quantum Science and Engineering, Huazhong University of Science and Technology, Wuhan 430074, China*

**Hao Wu -** *State Key Laboratory of Material Processing and Die & Mold Technology, School of Materials Science and Engineering, Huazhong University of Science and Technology, Wuhan 430074, China; Wuhan National High Magnetic Field Center and Institute for Quantum Science and Engineering, Huazhong University of Science and Technology, Wuhan 430074, China*

**Li Yang -** *State Key Laboratory of Material Processing and Die & Mold Technology, School of Materials Science and Engineering, Huazhong University of Science and Technology, Wuhan 430074, China; Wuhan National High Magnetic Field Center and Institute for Quantum Science and Engineering, Huazhong University of Science and Technology, Wuhan 430074, China*

**Wen Jin -** *State Key Laboratory of Material Processing and Die & Mold Technology, School of Materials Science and Engineering, Huazhong University of Science and Technology, Wuhan 430074, China; Wuhan National High Magnetic Field Center and Institute for Quantum Science and Engineering, Huazhong University of Science and Technology, Wuhan 430074, China*

**Bichen Xiao -** *State Key Laboratory of Material Processing and Die & Mold Technology, School of Materials Science and Engineering, Huazhong University of Science and Technology, Wuhan 430074, China; Wuhan National High Magnetic Field Center and Institute for Quantum Science and Engineering, Huazhong University of*

*Science and Technology, Wuhan 430074, China*

**Wenfeng Zhang** - *State Key Laboratory of Material Processing and Die & Mold Technology, School of Materials Science and Engineering, Huazhong University of Science and Technology, Wuhan 430074, China; Wuhan National High Magnetic Field Center and Institute for Quantum Science and Engineering, Huazhong University of Science and Technology, Wuhan 430074, China; Shenzhen R&D Center of Huazhong University of Science and Technology, Shenzhen 518000, China*

**Author Contributions**

H.C. supervised the project. H.C. and G.Z. designed the project. G.Z. and W.J. grew the single crystals. G.Z. fabricated the Hall devices. G.Z., H.W., L.Y., and B.X. did the characterizations and measurements. H.C., G.Z., and W.Z. analyzed the results. G.Z. and H.C. wrote the paper and all authors commented on it.

**Notes**

The authors declare no competing financial interest.

**ACKNOWLEDGMENTS**

This work is supported by the National Key Research and Development Program of China (Grant No. 2022YFE0134600), the National Natural Science Foundation of China (Grant No. 52272152, 61674063, and 62074061), the Interdisciplinary Research Program of Huazhong University of Science and Technology (Grant No.

2023JCYJ007), the Shenzhen Science and Technology Innovation Committee (Grant No. JCYJ20210324142010030 and JCYJ20230807143614031), and the Natural Science Foundation of Hubei Province, China (Grant No. 2022CFA031).

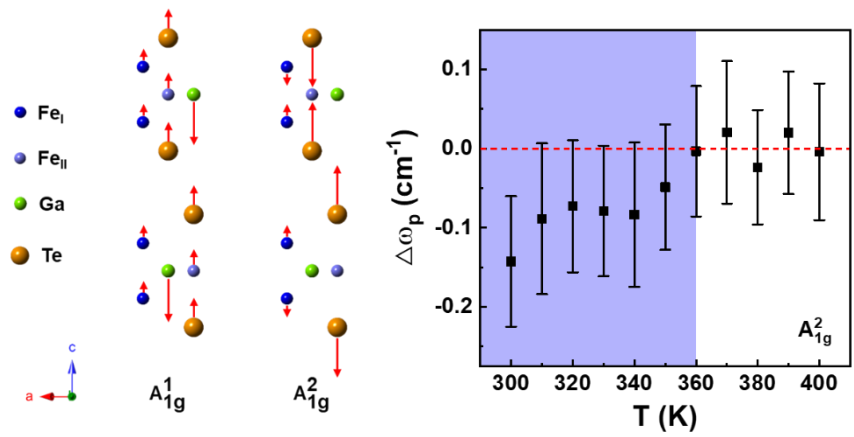

**For Table of Contents Only**

Supporting Information for

**Lattice Vibration, Raman Modes and Room-Temperature Spin-Phonon Coupling in Intrinsic 2D van der Waals Ferromagnetic $Fe_3GaTe_2$**

Gaojie Zhang[1,2], Hao Wu[1,2], Li Yang[1,2], Wen Jin[1,2], Bichen Xiao[1,2], Wenfeng Zhang[1,2,3], Haixin Chang[1,2,3,*]

[1]State Key Laboratory of Material Processing and Die & Mold Technology, School of Materials Science and Engineering, Huazhong University of Science and Technology, Wuhan 430074, China.

[2]Wuhan National High Magnetic Field Center and Institute for Quantum Science and Engineering, Huazhong University of Science and Technology, Wuhan 430074, China.

[3]Shenzhen R&D Center of Huazhong University of Science and Technology, Shenzhen 518000, China.

[*]Corresponding authors. E-mail: hxchang@hust.edu.cn

**This pdf file includes:**

1. Figure S1-S2.

2. Table S1.

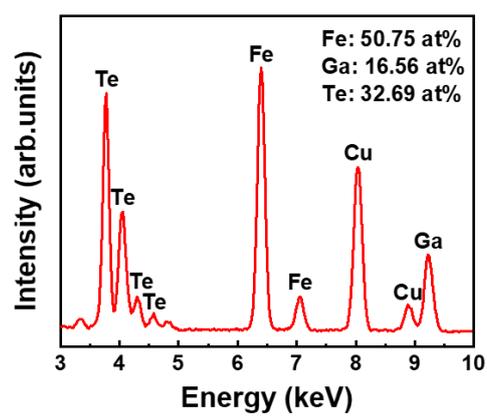

**Figure S1.** EDS spectra of an as-grown bulk $Fe_3GaTe_2$ crystal.

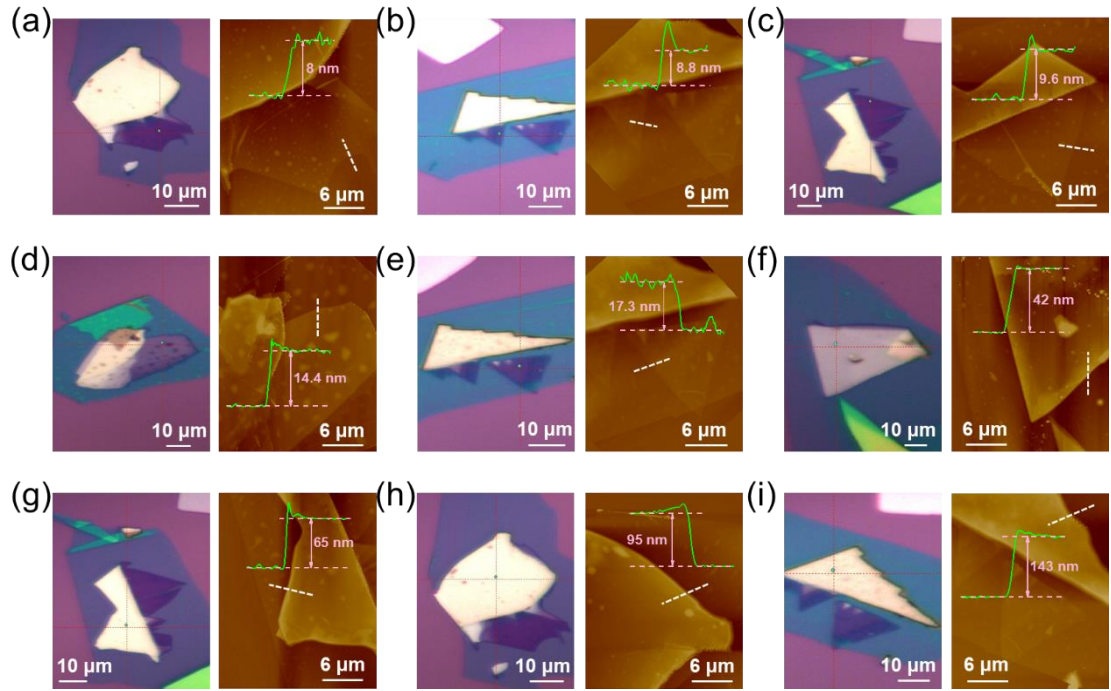

**Figure S2. Optical images and AFM tests of hBN-encapsulated Fe$_3$GaTe$_2$ nanosheets for room-temperature Raman measurements.** (a) 8 nm. (b) 8.8 nm. (c) 9.6 nm. (d) 14.4 nm. (e) 17.3 nm. (f) 42 nm. (g) 65 nm. (h) 95 nm. (i) 143 nm. The green points in optical images are the testing points. Specifically, Fe$_3$GaTe$_2$ nanosheets with different thicknesses are mechanically exfoliated from bulk crystals onto SiO$_2$/Si substrates. To avoid the oxidation and laser irradiation, a thin boron nitride (h-BN) flake is exfoliated and covered onto the Fe$_3$GaTe$_2$ nanosheet before Raman measurements using the dry transfer method in an Ar-filled glove box (H$_2$O, O$_2$ ≤ 0.1 ppm).

**Table S1.** In-plane lattice constants of bulk, bilayer, and monolayer $Fe_3GaTe_2$ atomic models.

| Thickness | a = b (Å) |
|---|---|
| Bulk | 4.074 |
| Bilayer | 4.071 |
| Monolayer | 4.067 |